\documentclass[twocolumn]{article}
\usepackage[usenames,dvipsnames]{color}
\newcommand{\ip}[2]{\langle #1 | #2\rangle}

\newcommand{\qs}[1]{| #1\rangle}
\title{Impossibility of comparing and sorting quantum states}
\author{A. John Arul
\\{\it Reactor Physics Division, Indira Gandhi Centre for Atomic Research}
\\{\it Kalpakkam 603 102, India}
\date{(July 17, 2001)}}
\begin{document}

\maketitle

\begin{abstract}
\noindent
{\bf Is there any point of principle that prohibits us from doing one or more forms of quantum information processing? It is now well known that an unknown quantum state can neither be copied nor deleted perfectly\cite{nqc,nqd}. Given a set of states which are not necessarily orthogonal, is it possible to compare any two states from the set, given some reasonable ordering of such states? Is it possible to sort them in some specific order? In the context of quantum computation, it is shown here that there is no quantum circuit implementing a unitary transformation, for comparing and sorting an unrestricted set of quantum states.}
\end{abstract}\\
\noindent
\vspace{3mm}

Comparing and sorting are fundamental computational process like copying, deleting, storing etc. With recent advances in quantum information processing\cite{deutch, qcomp} it is pertinent question to ask what computations are exactly permissible with a quantum model of computation compared to classical models of computation. For instance, in quantum computation while {\it permutation} and {\it swap}\cite{qcomp} operations of non-orthogonal states are allowed, unknown quantum states cannot be {\it copied} perfectly\cite{nqc,qcomp}, unless one chooses to work with, orthogonal set of quatum states. It was also shown recently \cite{nqd}, that a particular form of {\it deletion} is  not allowed in the realm of quantum computation. The natural question to ask is, can one compare two non-orthogonal states using a quantum computer implementing some unitary transformation? Let ${\mathcal S}=\lbrace \qs{\psi_k}\rbrace(k=1,2..N)$ be some ordered set of quantum states defined such that $\qs{\psi_i} < \qs{\psi_j} $ iff $i < j$ and $\qs{\psi}$ are elements of Hilbert space ${\mathcal H}$ of dimension $D$. The comparator transformation $U:$ ${\mathcal H^{'}} \rightarrow {\mathcal H^{'}}$ is defined as
\begin{eqnarray}
\qs{\psi_i}\qs{\psi_j}\qs{\Sigma}\rightarrow \qs{1}\qs{\Sigma^{'}} \nonumber \\
\qs{\psi_j}\qs{\psi_i}\qs{\Sigma}\rightarrow \qs{0}\qs{\Sigma^{''}}  \label{eq1}
\end{eqnarray}
\noindent
where $\qs{0}$ and $\qs{1}$ are orthogonal basis states, and $\qs{\Sigma} $ is ancilla state. $\qs{\Sigma^{'}}$ and $\qs{\Sigma^{''}}$ are combination of ancilla and inputs after transformation. By looking at the first state (on rhs of equation~\ref{eq1} ) after the transformation, we should be able to infer whether the input state one is less than the second (on lhs of equation~\ref{eq1} ). The unitarity of the comparator transformation implies
\begin{equation}
\ip{\psi_j}{\psi_i}\ip{\psi_i}{\psi_j} \quad=\quad \bf{0}
\end{equation}
which is true only if $\qs{\psi_i}$ and $\qs{\psi_j}$ are orthogonal.
The inference is, non-orthogonal states can not be compared using a quantum comparator. In a classical model of computation, elements can be sorted by binary comparison \cite{ccomp}, for example {\it bubble sort, merge sort, quick sort} etc, to name a few. As is well known, in classical computation one can sort even without doing comparison between the input elements, if we have some prior information about the elements, using {\it count sort} or {\it bucket sort}~\cite{ccomp}. This leads us to the question whether quantum states can be sorted using a quantum circuit, although we cannot compare them? Let us see if the unrestricted collection of states ${\mathcal S}$ can be sorted into some specified order. Here, by {\it unrestricted} collection of states we mean that the states need not form a mutually orthogonal set; they need not even form a linearly independent set.  The only requirement on the set ${\mathcal S}$ is that its elements can be given a certain total ordering. The orthonormal basis set spanning the Hilbert space ${\mathcal H}$ can be given an order in terms of the binary values they represent. For example, in the two dimentional case one can impose that $v(\qs{0})=0$ and $v(\qs{1})=1$. Given this, linear superpositions can be assigned a value equal to the  expectation value. i.e., $v(\alpha \qs{0}+\beta \qs{1})=0\! |\alpha|^{2}+1\! |\beta|^{2}$. Here $v(.)$ is a mapping from elements of ${\mathcal H}$ to ${\mathcal {R^{+}}}$. The sorting transformation is defined as

\begin{equation}
\qs{\psi_q}\qs{\psi_i}...\qs{\psi_j} \qs{\Sigma}\rightarrow \qs{\psi_i}\qs{\psi_j}...\qs{\psi_q}\qs{\Sigma^{'}} 
\end{equation}

\noindent
where $i< j < q$. That is, the output sequence is a monotonically increasing one for every input sequence. For three specific states $\qs{\psi_1},\qs{\psi_2},\qs{\psi_3} $, some of which are not orthogonal, 

\begin{eqnarray}
\qs{\psi_1}\qs{\psi_2}\qs{\Sigma}\rightarrow \qs{\psi_1}\qs{\psi_2}\qs{\Sigma^{'}} \nonumber\\ 
\qs{\psi_3}\qs{\psi_2}\qs{\Sigma}\rightarrow \qs{\psi_2}\qs{\psi_3}\qs{\Sigma^{''}}\label{eqsort}
\end{eqnarray}
once again by the unitarity of the transformation we require,
\begin{equation}
\ip{\psi_3}{\psi_1} \quad=\quad \ip{\psi_2}{\psi_1}\ip{\psi_3}{\psi_2}\ip{\Sigma^{''}}{\Sigma^{'}}\label{eqsortr}
\end{equation}

\noindent
If $\qs{\psi_2}$ and $\qs{\psi_3}$ are assumed to be orthogonal, then $\ip{\psi_2}{\psi_3} =0 $, implying $\ip{\psi_3}{\psi_1}=0$
which need not be true. This means a set of quantum states which do not necessarily form an orthogonal set (some are not orthogonal) {\it cannot be sorted} by a unitary machine. 

Important operations such as comparing and sorting of an unrestricted set of quantum states turnout to be impossible in a quantum computer.  However, it should be noted that one can work with an orthogonal set of states, and do operations like comparison and sorting, as one can also copy such orthogonal states.  For instance Grover's database search algorithm\cite{grover} works on orthogonal set of states. This result is intimately connected to the nature of quantum information contained in a quantum state, all of which is not accessible to direct manipulation ~\cite{qcomp}.

\noindent
{\bf Acknowledgements}\\ 
I thank V. Sridhar, MSL for useful disucussions and comments. 
\vspace{5mm}\\
Correspondence and request for materials should be addressed to the author.
\begin{verbatim}
email: arul@igcar.ernet.in
\end{verbatim}

\end{document}